# The Criteria for Interfacial Electro-Thermal Equilibrium

Albert E. Seaver

Electrostatics Consultant
7861 Somerset Ct.
Woodbury, MN 55125-2333, USA
e-mail: aseaver@ electrostatics.us

*Abstract*—When the surface of a first material is brought into contact with the surface of a second material the contact region is called an interface. Since the time of James Clerk Maxwell it has been customary to treat a material electrically as having well-defined bulk properties and having surfaces of zero-thickness. From this point of view when two surfaces come together the interface also has zero thickness. However, in practice, an electrical potential difference is found across many interfaces, and the assumption of a zero-thickness interface leads to problems when attempting to describe this phenomenon at an interface. To get around the problem, it is customary to assume a potential at some boundary and then compute an effective thickness of the interface. In the simplest model the potential is assumed to be due to charge separation at the interface, and the potential difference occurs across the separation distance. The interface is then treated as a capacitor layer with its plate separation or interfacial thickness referred to as the electrical double layer thickness. In other models this interfacial thickness is known as either the depletion region or the Debye length. Often the interfacial thickness is estimated to be on the order of nanometers. With the advent of nanotechnology at least one dimension of the nano-material is in the nanometer range and suddenly the interface thickness, if it exists, can be a substantial fraction of the bulk material thickness of that nano-material. Besides nanotechnology the interface and interfacial potentials are important in areas such as neurological and biological systems, in triboelectric (contact) charging and in thermoelectric generators/refrigerators. In order to obtain a better understanding of the interface this paper reviews Maxwell's original argument to justify a zero-thickness-surface and reexamines the interface problem assuming electrical charges are actually particles having a finite thickness. Thermodynamics requires that in thermal equilibrium any movement of free charge cannot produce a net electrical current anywhere in the materials or across their interface. For materials in contact and in thermal equilibrium this reexamination gives a set of equations that can be called the interfacial electro-thermal equilibrium (IETE) criteria. From these a well-defined interfacial potential results.

I. INTRODUCTION

The interface – the region of contact between two materials – is ubiquitous, i.e., it can be found almost everywhere. The interface is important in many electrical phenomena, such as triboelectric or contact charging [1], in electronic devices such as the junction diode [2,3,4], transistor [3,4], thermocouple [5], thermoelectric generator [6,7] and thermoelectric refrigerator [7]. Furthermore, the interface plays an important part in both neurological and biological systems through electrical signal transport in the nervous system and material transport through cell membranes, respectively [8,9]. The interface is even important within many materials, as for example, dielectrics when viewed on a microscopic scale are rarely homogeneous – their composition being a mixture of amorphous and crystalline regions [10], and it has been reported that these interface regions are host to hopping sites [11].

The microelectronics industry evolved from the transistor – a three-layer, two-interface device. The transistor gave way to the microchip – a device containing several transistors and circuits – which is created by depositing micrometer sized layers of materials that make up the multi-interfaced chip. In turn, the microchip gave way to the very large scale integration (VLSI) chip on which millions of transistors and circuits are deposited. As the layers of materials deposited to make up the chip have become thinner, layer thickness is no longer being measured in micrometers (1 micrometer = $10^{-6}$ m) but rather in nanometers (1 nanometer = $10^{-9}$ m). As a result the microelectronics industry and the acronym MEMS for micro-electro-mechanical systems is becoming the nanoelectronics industry with the acronym NEMS referring to nano-electro-mechanical system [12]. Today nano-layered electronic, photonic and material devices consisting of repeating layers of two different materials – where the thickness of the layer is defined in the nanometer range – offer great promise because their properties tend to differ from that of similar devices whose layer thickness is in the micrometer range or above [12,13]. As a general rule most manufactured nanoelectronic interfaces today are either solid-nanosolid-solid, solid-nanosolid-liquid, solid-nanosolid-gas or repeating nanosolid interfaces.

There are a lot of similarities between the interfaces encountered in the nanoelectronics industry and in the interfaces found in nature in neurological systems. For example, in the nervous system a nerve is a bundle of neurons, and a neuron consists of three parts; a cell body from which extend one or more fiber-like axons, and from which also extend one or more short-fiber-like dendrites [8]. The axon of one neuron is typically very close to the dendrite of another neuron and electrical interaction occurs at this axon-dendrite gap called the synapse, which typically is about a 20 nanometers gap of fluid ([9] see Fig. 9.3, and Fig. 9.7). In essence neurological signals are transmitted through these neurological nanometer-gaped synapse devices where the synapse is simply a two-interface device. Neurological interfaces tend to be predominantly solid-liquid interfaces.

Similar arguments can be extended to biological systems. Biological cells come in all sizes, however, most cells are in the 500 – 40,000 nanometer range and the cell membrane – which surrounds the cell – is only about 8 nanometers thick ([9] pp. 12-13). Here again the cell membrane is simply a two-interface device. Biological interfaces tend to be predominantly solid-liquid or liquid-liquid interfaces.

Electrical interactions occur at an interface, and generally speaking, the analysis differs depending on if the interface is a solid-solid, solid-liquid or liquid-liquid interface. Solid-gas and liquid-gas interfaces are usually referred to simply as surfaces. When first attempting to analyze any of these interfacial situations the two materials brought into contact are usually assumed to have uniform properties and the interface is assumed to be of zero-thickness and simply considered as a boundary condition. However, there is much scientific evidence suggesting that an electrical potential difference occurs across most of these interfaces. To explain such a potential difference a further argument must be made such as the development of a depletion region around a solid-solid semiconductor interface [2], or a charged double layer with a double layer thickness region around a solid-solid or solid-liquid interface [14,15,16], or a Debye separation length near a solid-liquid or a liquid-liquid interface [16]. As a result every interface has a potential and an associated length (depletion layer depth, double layer thickness, Debye length, etc.). It has been argued that the properties vary across the interface and it is the interaction of short and long-range forces that determine the profile of the interface (see [12] p.740). In any event, potentials in the millivolt range and interface thickness in the nanometer range are found with many of these interfaces.

To summarize, the onset of nanotechnology and nanoelectronics has resulted in a renewed interest in the interfaces, and the desire for a more in-depth understanding in both neuroscience and bioscience has resulted in a renewed interest in the interface. It therefore seems worthwhile to undertake a study that might allow a comprehensive look at the general electrical interaction at an interface. It would appear to be extremely beneficial if a single electrical-criteria could be mathematically developed that would be applicable to any interface be it a solid-solid, solid-liquid or liquid-liquid interface, or be it a common surface either solid-gas or liquid-gas. Such an undertaking is the basis of this paper.

## II. Outline

In order to reach the goal of determining a single set of mathematical criteria at an interface this paper first reviews the 130-year-old electrical argument that lead to the zero thickness surface assumption and then reviews the thermal equilibrium requirement of thermodynamics, which requires an ohmic material to have a zero electric field everywhere inside its volume. Then, the zero-thickness assumption is reexamined, and – based on a more modern day un-derstanding of electricity – this zero-thickness interface assumption is removed and the physical and mathematical situation of an interface is reexamined. It is argued that, when free charges are present in either or both materials, thermodynamics actually requires an electric field to be set up in both materials in the region of the interface. This electric field must produce conduction current to exactly compensate for the diffusion current that naturally occurs due to the difference in the number density of free charges in the interface region. Specifically, two opposite flowing currents are found to exist in the interface region, diffusion current and conduction current, but in thermal equilibrium the net current at any point is always zero.

### III. ZERO THICKNESS SURFACE ARGUMENT

James Clerk Maxwell (1831-1879) gave the world a greatly improved understanding of electricity, magnetism and electromagnetic waves with the publication in 1873 of his famous book *A Treatise on Electricity and Magnetism* [17,18]. His feat was even more impressive, since during Maxwell's time there was a big debate as to the proper way to describe electrical interactions ([1] pp. 4-5). One camp, primarily in Germany, believed in "action at a distance" to described the gravitation, electric and magnetic forces. In this camp, also referred to as the mathematician's camp, forces are most important and potentials much less important. The other camp, primarily in England, believed in Michael Faraday (1791-1867) and his view that lines of force extended through all space. The Faraday camp considered the potential – a quantity that satisfies a certain partial differential equation – as most important. Maxwell considered this split in belief between the Faraday camp and the mathematician camp to be of such grave consequence that he addressed the issue in the preface of his first edition ([17] pp. vii - xi). It was a credit to Maxwell's genius that he was able to convincingly argue a common connection between these two camps within his book.

Although the above-mentioned split into "lines of force" and "action at a distance" camps was a major deterrent to progress during Maxwell's time, another split regarding how to treat electricity at a surface also existed. One camp believed electricity on a surface could be described as being located at points on a two-dimensional surface, such that the surface had zero thickness. The other camp argued that a surface had to have a finite thickness in order for electricity to inhabit the surface. As little was know about the makeup of electricity at that time, this split into "zero-thickness-surface" and "finite-thickness-surface" camps was also hindering advancements in the study of electricity.

*A. Zero thickness surface*

Maxwell addressed the two camps regarding the thickness split in his book ([17] p. 72, Sect. 64) and chose to define the electric volume density $\rho$ at a given point in space as

> "the limiting ratio of the quantity of electricity within a sphere whose center is the given point to the volume of the sphere, when the radius is diminished without limit."

However, he noted that if electricity were confined to a surface, then the electric surface density $\sigma_s$ of a point on the surface, if defined according to the method given above, would be infinite. {Note, in his book Maxwell used $\sigma$ to define the surface charge density, but in the discussion here it will be listed as $\sigma_s$ whereas $\sigma$ will be reserved for electrical conductivity.} Instead Maxwell defined the electric surface density $\sigma_s$ at a given point on a surface as

> "the limiting ratio of the quantity of electricity within a sphere whose center is the given point to the area of the surface contained within the sphere, when the radius is diminished without limit."

These definitions are consistent with the definitions of the volume charge density $\rho$ and surface charge density $\sigma_s$ that are in use today. The "diminished without limit" statement of Maxwell was not accepted without reservation and for the next century textbooks cautioned that the limit must not decrease beyond a finite but small limit in which, for example, the volume still contained a reasonable number of atoms [19]. Over time this precaution has been dropped and in modern textbooks the definitions of $\rho$ and $\sigma_s$ are simply stated or their equations written without further discussion [20,21].

Maxwell then joined the "zero-thickness-surface" camp as discussed below.

*B. Finite thickness surface*

Although Maxwell joined the "zero-thickness-surface" camp he knew he had to address the point of view of the finite-thickness camp so he wrote ([17] p. 72, Sect. 64)

> "Those writers who supposed electricity to be a material fluid or a collection of particles, were obliged in this case to suppose the electricity distributed on the surface in the form of a stratum of a certain thickness $\theta$, its density being $\rho_0$, or that value of $\rho$ which would result from the particles having the closest contact of which they are capable. It is a manifestation of this theory [that] $\rho_0 \theta = \sigma_s$."

Maxwell went on to justify not using the finite thickness method by writing:

> "**There is, however, no experimental evidence** either of the electric stratum having any thickness, or **of electricity being a fluid or a collection of particles**."

So why did Maxwell not accept what is basically accepted today, namely, that electricity is the movement of charged particles of finite size? Ancient

Greek philosophers Leucippus (ca. 480-420 BC) and his student Democritus (460-370 BC) are believed to be the first to predicted the existence of the atom and they named the particle "atomos," meaning "uncut or indivisible" [22], but the concept languished until 1803 when John Dalton (1766-1844) proposed a systematic set of postulates to describe the atom [23]. Later, Maxwell is credited with developing the idea mathematically in the kinetic theory of gases [24]. Although the atom was initially thought to be an indivisible, indestructible, tiny ball, by 1850 evidence was accumulating that the atom was itself composed of smaller particles [25]. However, the historical timeline of discovery shows that it was not until near the time of Maxwell's death that a connection between atoms and charges was being considered [26].

### C. Did Maxwell change his mind?

It is important to remember that Maxwell's untimely death occurred in 1879 and it was not until 1897 that J. J. Thompson (1856 – 1940) discovered the electron. During Maxwell's time electricity was considered a weightless (i.e, massless) fluid ([1] p. 2) and there was no knowledge of the existence of the electron and proton ([1] p. 5). However, Maxwell had accepted gases as being composed of particles and had also been developing the kinetic theory of gases, so clearly, in his latter years he had become enamored with the concept of particles. Hence, the question becomes, by the time of his death, did Maxwell suspect electricity was in reality related to a collection of charged particles? There is some compelling evidence that Maxwell may have recognized this before he died. After Maxwell's death, the mathematician William D. Niven (1843-1917) was asked to edit the second edition of Maxwell's book on Electricity and Magnetism. Niven noted in the preface to the second edition that Maxwell ([17] pp. xii-xiv)

> "contemplated considerable changes: viz the mathematical theory of the conduction of electricity in a network of wires..."

Unfortunately, after reviewing Maxwell's notes Niven also wrote that he (Niven) had not found himself in a position to add anything substantial to the work as it stood in the former edition. One can only speculate on just what "considerable changes" Maxwell would have made to the "mathematical theory of the conduction of electricity in a network of wires" had he lived just a little longer.

Certainly the particle nature of electricity would suggest that the definition of a surface of zero thickness would be impossible if particles of finite thickness occupied it. Furthermore, a "network of wires" implies connection between wires, so Maxwell would need to have readdressed the wire interface.

### D. The situation at an interface

The simplest "network of wires" would be two wires connected together, and Maxwell had already noted the Peltier effect – namely, when a current of electricity crosses the junction of two metals, the junction is heated when the cur-rent is in one direction and cooled when it is in the other direction ([17] pp. 368-369, Sect. 249).

Likewise, Maxwell had also noted the Seebeck effect – namely, thermoelectric currents in circuits of different metals with their junctions at different temperatures, showed junction potentials which did not always balance each other in a complete circuit ([17] pp. 370-371, Sect. 249).

Maxwell further noted that the thermoelectric current (Seebeck effect) must disappear if a circuit consisting of wires connected in series is at a uniform temperature. Otherwise, Maxwell noted ([17] p. 370, Sect. 250),

> "…there would be a current formed in the circuit, and this current might be employed to work a machine or to generate heat in the circuit, that is, to do work, while at the same time there is no expenditure of energy, as the circuit is all at the same temperature, and no chemical or other change takes place."

Such a current (or current density **J**) would violate both the first and second laws of thermodynamics [27]. In simple terms the first law states a conservation of energy, so there must be an energy source in order to obtain useful work, whereas the second law states for useful work to be extracted from heat a temperature difference must exist. In essence, Maxwell gave the standard thermodynamic argument that **J** = 0 in thermal equilibrium.

In what follows this paper reexamines the situation Maxwell addressed nearly 130 years ago and then adds to it a more complete discussion based on electricity being the result of particles of *finite* dimension.

## IV. STANDARD ANALYSIS

Whenever a new point of view is presented, it is always wise to first present the old point of view in sufficient detail that an easy comparison can be made. This insures that the arguments for the new point of view are self-contained.

Although this paper is concerned with the interface between any two materials (solid, liquid or gas) much of the development will be described by considering the simple situation of a single solid conductor in contact with another solid conductor. As will be discussed this eliminates much of the complications without loosing the procedure used to develop the general criteria for interfacial electro-thermal equilibrium (IETE).

### A. Thermodynamic Considerations

In thermodynamics a *simple* system is described as macroscopically homogeneous, isotropic, uncharged and chemically inert and is sufficiently large that surface effects can be neglected, and, furthermore, the system is not acted on by external electric, magnetic or gravitational fields [28]. If an *adiabatic* wall – a wall that restricts heat flow – is placed around a system then the system is defined as *closed* ([28], p. 15). A single wire is normally analyzed in a simple closed system, but – as will be discussed later in Section V.A – even a single wire cannot be considered a simple system even if a perfect vacuum surrounds it. Furthermore, two dissimilar wires in contact cannot be considered a simple system because when in contact the wires are not a homogeneous system unless they are of identical composition.

### B. Electrical charges

The first postulate of thermodynamics is of great interest in this paper; namely, ([28], p. 12) ) "There exists particular states (called equilibrium states) of simple systems that, macroscopically, are characterized completely by the internal energy $U$, the volume $V$ and the mole numbers $N_1, N_2,…$ of the chemical components."

However, in studying electrical systems the interest is not so much in the chemical components as it is in free electrons and the ability of the members of each chemical component to give up (detach) or accept (attach) a free electron. Not all members of a chemical component have a free electron, which is defined as an electron that has the ability to freely move when under the action of a force. When a free electron moves from a member of a chemical component it makes that member positively charged, and when a free electron attaches to a member of any chemical component it makes that member negatively charged. In a solid when acted on by a force the free electrons are free to move, but in a liquid or a gas the free electrons as well as the positively charged and negatively charged members of the chemical components are free to move. When any of these free charges are acted on by any gradient a charge flux or charge current density **J** is produced ([28] pp. 293-296).

### C. Conductors, Semiconductors and Insulators

For any material if the number of the $i^{th}$ chemical component is $N_i$ then, at a given temperature $T$ the number of free charges $n_i$ associated with $N_i$ is $f_i N_i$, hence, $f_i = n_i/N_i$ is the fraction of the number $N_i$ that have a free charge associated with the $i^{th}$ chemical component. From this information the distinction between conductors, semiconductors and insulators can be easily defined. If $f_i \ll 1$ for all $i$ then a material is a good insulator. If $f_i \sim 1$ for some $i$ and $f_j \ll 1$ for all $j \neq i$ while $N_i < N_j$ for some $j$ then the material is a semiconductor. Finally, if $f_i \sim 1$ for some $i$ and $f_j \ll 1$ for all $j \neq i$ while $N_i > N_j$ for all $j \neq i$ then the material is a good conductor. The electrical conductivity $\sigma_i$ is a property of the $i^{th}$ chemical component indicative of its ability to transport its free charges when acted on by an electrical force. The electrical conductivity $\sigma = \Sigma \sigma_i$ is a property of the material indicative of its ability to transport all its free charges when acted on by an electrical force.

### D. Ohm's Law

In 1827 Georg Simon Ohm (1789-1854) published his famous mathematical relationship between the voltage $V$ and current $I$ of a homogeneous conductor having a property $\sigma$, which offered a resistance $R$ to the flow of current. This relationship, known as Ohm's law is most often written as *V* = *IR* [29]. The resistance *R* depends on the shape of the conductor, and for any infinitesimally thin slab along the conductor the resistance is directly proportional to the thickness, inversely proportional to its cross-sectional area *A*, and inversely proportional to its electrical conductivity $\sigma$. It has been recognized that the current per unit area, i.e., the charge flux or electric current density **J**, is controlled by the electric field **E** = -$\nabla V$ in a conductor. As a result an alternative relationship, known as the field form of Ohm's law is given by [29]

$$\mathbf{J} = \sigma \mathbf{E}. \qquad (1)$$

As Maxwell astutely noted Ohm's law is only applicable to *homogeneous* conductors ([17] p. 362).

### E. Thought Experiment #1 – One Conductor

The simplest electrical system is a conductive wire. For a simple closed system consisting of a wire in thermal equilibrium thermodynamics requires **J** = 0; otherwise, a part of the current could be used to produce work. If **J** = 0, then from Ohm's law (1) it can be concluded that the electric field **E** must be equal to zero, since $\sigma$ is finite. It is the above thermodynamic analysis that produces the following well-known conclusion:

**Conclusion #1** (Requires Ohm's law and **J** = 0)
The electric field **E** must be equal to zero *everywhere* inside an *isolated* good conductor in thermal equilibrium.

### F. Thought Experiment #2 – One conductor with excess charges

If an excess of free charges are now uniformly placed inside the conductor (now no longer definable as a simple system), the standard argument is that Coulomb repulsion forces will cause the charges to move away from each other, eventually moving them to the outer surfaces of the conductor. In thermal equilibrium the electric field **E** inside the conductor must be zero (based on Thought Experiment #1), so the charges on the surface must also rearrange themselves to insure **E** is zero everywhere within the conductor. As a result for the situation of an isolated conductor with excess free charge, the surface indirectly becomes important and it must be analyzed further. The standard method is to apply Gauss's law just outside the charged conductor from which it can be concluded that the charges reside inside the Gaussian surface, and then apply Gauss's law just inside the conductor surface where no electric field exists and hence no charges are enclosed from which it can be concluded that the charges must reside on the surface [29]. Since the two Gaussian surfaces can be infinitesimally close to each other the conductor surface can be of zero thickness. Thus, a second conclusion appears evident:

**Conclusion #2a** (Requires Gauss's law and Conclusion #1)
Electrically a surface can have zero thickness.

### G. Point charges easily satisfy the zero thickness surface

If $\mathbf{E} = 0$ everywhere inside the conductor then the charges, if they have a finite size, must reside outside the conductor. Since there is another medium outside the conductor (even if it is free space), this would imply that the charges exist in the other medium. However, if the charges are considered as point charges, then they can have zero thickness and can reside on a surface defined to have zero thickness. As a result, the charges, if they can be defined as point charges, exist only at the interface and are not located inside either the conductor or the medium surrounding the conductor. Thus, a further argument for a zero-thickness surface exist, namely:

> **Conclusion #2b** (Requires point charges and Conclusion #1)
> Electrically a surface can have zero thickness.

### H. Thought Experiment #3 - Two conductors

A reasonably uncomplicated electrical system consists of two conductive wires in contact with each other. This is an "almost" closed system where each wire is homogeneous, but the combination connected together is not, so it violates one of the requirements of a closed system. For this two-wires-connected system in thermal equilibrium thermodynamics require $\mathbf{J} = 0$, otherwise, a part of the current could be used to produce work. If $\mathbf{J} = 0$ everywhere, then from Ohm's law (1) it can be concluded that the electric field $\mathbf{E}$ must be equal to zero *everywhere* inside *each* conductor and therefore inside both conductors when in thermal equilibrium. As a result, if charges exist at an interface, they must be point charges and reside on a surface of zero thickness, because otherwise they would be in one of the conductors and then, the electric field would not be zero in that conductor. Thus, a further argument for a zero-thickness surface exist, namely:

> **Conclusion #3** (Requires point charges and Conclusion #1)
> Electrically an interface must have zero thickness.

## V. NEW ANALYSIS

It can be argued that the above analysis – based on Ohm's law and thermodynamics – has withstood the test of time because many problems in electrical engineering have been solved using Conclusion #1, namely, "the electric field everywhere inside a good conductor is zero." It would therefore seem ludicrous to question any of the above analysis.

However, there are questions and some real problems still unsolved with regard to certain aspects of electrical phenomena at an interface. The first and foremost question is why in thermal equilibrium does a potential difference exist across so many interfaces? This is not a trivial question, and its consequences are of great importance. For one example, triboelectric (contact) charging is not predicted by the above analysis when it is extended to semiconductors and insulators and as a result the cause of triboelectric charging is still not really understood [1,30]. For another example, the properties of materials in nanotechnology, which reside on a surface, have at times been found to differ greatly from the same materials in the bulk state [12,13]. Based on these examples as well as other areas in neurological and biological systems, which were briefly described in Section I above, it seems prudent to "review" the standard interface analysis in case something has been overlooked that could shed further understanding of potentials and their influences at interfaces.

### A. Real charges have a thickness

An electron has a diameter of about $10^{-14}$ m [31]. In a solid conductor it is the movement of free electrons that constitutes a current. However, for the general case a material can be a solid, liquid or gas and the electrons that are considered free to move under the action of a force can, for a liquid or a gas, attach to any of the members of the $N_i$ chemical components leaving some fraction(s) of the member(s) negatively charged. Likewise, if a free electron moves from a member of its chemical component it will leave that member positively charged and that positively charged member (in liquids and gases) would also be free to move under the action of a force field. Since all the members of any chemical component are of finite size, when they become charged, these charged members are also of finite size. However, restricting the discussion back to conductors for the moment, if these charges are of finite size, and must reside outside the conductor so that **E** = 0 everywhere inside the conductor, then the charges must exist in the *medium* just outside the conductor. If the conductor is situated in free space (i.e., space void of any chemical components) then the excess charges would be forced to "invade" the free space for a conductor system in thermal equilibrium. In other words even in Thought Experiment 1 and Thought Experiment 2 the system is not a simple conductor, but rather a simple conductor surrounded by another material, even if that material is free space. The only thing a perfect vacuum or free space offers is the absence of a collision mechanism to entice transport of a free electron at the surface out into the free space.

A further problem exists for conductive solids when some free electrons are removed to make the solid positively charged. For example, for a single chemical component material, the remaining free electrons in the solid must adjust themselves so that the surface becomes positively charged. However, this positive charge does not reside on the surface, but rather on the outermost members of the single component material. In other words if charges have a finite thickness, the concept of a zero thickness surface becomes physically impossible.

To summarize, if material A is placed against material B, where material A is a solid or liquid and material B is a solid, liquid, gas or vacuum, then at the minimum the thickness of a real interface (material B being either solid or liquid) or of a real surface (material B being either a gas or vacuum) must consist of at least some portion of the outermost volume element of material A and at least some portion of the outermost volume element of material B, the limit not decreasing below some thickness $\theta_A$ in material A and $\theta_B$ in material B as described by Maxwell ([17] see p. 72, Sect. 64) and presented in Section III.B above.

From the forgoing it can be concluded that if charges are of finite thickness then logical arguments for a finite-thickness surface and a finite-thickness interface exist, namely:

>**Revised Conclusion #2** (Requires charges of finite dimension)
>A surface must have a finite thickness.

And

>**Revised Conclusion #3** (Requires charges of finite dimension)
>An interface must have a finite thickness.

### B. Two conductors in contact: System is no longer homogeneous

The thesis of this paper is tied up in the following statement. Because free charges can move across boundaries, when two homogeneous conductors are in contact it is wrong to think of the system as two individual homogeneous conductors, but rather it is appropriate to consider the system as one inhomogeneous conductor. By extension, the same argument holds true for any two materials in contact even if only one type of free charges, namely, free electrons can move across the interface, the other types of charges being constrained by further restrictions such as, for example, impermeability.

The following argument should hold true near the interface of any two materials. For simplicity, consider two one-component-metal solid conductors; conductor A on the left of the interface and conductor B on the right with the normal of the interfacial plane in the $x$ direction (which for discussion is the horizontal direction). Before being connected together A and B were both charge neutral. In A the number of free electrons (per unit volume) $n_A$ is defined by some fraction of the total number of chemical component metal atoms (per unit volume) in A. By the same type of argument $n_B$ is the number of free electrons (per unit volume) in B. If, in conductor A, a free electron happens to be located next to the interface between the two conductors, and if $n_A > n_B$, then, since a free electron is a free electron no matter which conductor it happens to be in, there is a driving force on the free electrons in A proportional to the gradient $-dn_A/dx$ which will cause diffusion of free electrons in conductor A across the interface into conductor B. This gradient drives free electrons from A to B. On the other side of the interface is a driving force on the free electrons in B proportional to $-dn_B/dx$, which will cause further diffusion into B of the free electrons that came from conductor A. As a result, many free electrons in A will attempt to drain (diffuse) into B. For a diffusion constant $D$ and charge density $\rho = q_e n$ where $q_e$ is the charge of the electron, the result will be a diffusion current $\mathbf{J}_D = -D\nabla\rho$ from A to B in the regions on both sides of the interface.

However, as these free charges (electrons) move from A to B they make re-gion A near the interface positive and region B near the interface negative. As a result an electric field **E** is set up in the region of the interface which gives a conduction current $\mathbf{J}_\sigma = \sigma\mathbf{E}$ and this current drives free electrons back from B to A. These two currents will establish themselves in such a way that for every electron that diffuses from A to B there will be another free electron that will move from B to A due to the action of the electric field. At equilibrium the net result is the total current density – the sum of the conduction current and diffusion current – will be zero at every point within A and B in the entire region around the interface. At some distance from the interface in both A and B the charge density gradient will go to zero at which point the electric field will also no longer exist and the remaining length of the conductors will be satisfied by the condition that the electric field is zero.

The criteria for two isolated but connected conductors in thermal equilibrium becomes

$$\mathbf{J} = \sigma\mathbf{E} - D\nabla\rho \qquad (2)$$

where, near the interface, a charge density gradient $\nabla\rho$ exists and further away from the interface this gradient goes to zero and there (2) reduces to (1).

It is the above thermodynamic analysis that produces the new conclusion of this paper, namely:

> **Revised Conclusion #1** – Conductors only
> (Requires charge transport equation {see (2)} and $\mathbf{J} = 0$)
>
> The electric field **E** must be equal to zero *everywhere* inside good conductors in thermal equilibrium – except near an interface where an electric field must exist to counter the diffusion of free charges across the interfacial boundary. The current density on the other hand must be zero everywhere in order to satisfy the laws of thermodynamics.

This is a new way of looking at an interface, but it requires the recognition that an interface does not obey Ohm's law and that it is incorrect to apply the old Conclusion #1.

### C. Extension to any material from conductor to insulator

It is well known that most problems in electrostatics do not follow Ohm's law and the full equation of charge transport must be used [32]. Extension of the criteria to any materials be they solids, liquids, or gases and be they classified as conductors, semiconductors or insulators can be done by realizing that in the absence of a magnetic field– the charge transport equation of any material is given by [33]

$$\mathbf{J} = \rho\mathbf{v}_{d0} + \sigma\mathbf{E} - \sum_i D_i\nabla\rho_i - \nabla T\sum_i G_i\rho_i \;. \qquad (3)$$

where the material flows with drift velocity $\mathbf{v}_{d0}$, where the summation is over all charged species, and where the $i^{th}$ specie has a charge density $\rho_i$, an ordinary translational diffusion coefficient $D_i$ and a thermal diffusion coefficient $G_i$. At times $D_i$ is referred to as the diffusion constant and $G_i$ is also called the thermophoresis coefficient. The summation is written because each of the species has its own transport equation [34]. As a result, in (3) both $\rho$ and $\sigma$ are summations over all species.

If the material is at a constant temperature ($\nabla T = 0$) and if there is no material flow ($\mathbf{v}_{d0} = 0$), then, (3) reduces to

$$\mathbf{J} = \mathbf{E}\sum_i \sigma_i - \sum_i D_i \nabla \rho_i . \qquad (4)$$

When only one charged species dominates (4) reduces to (2) but for the general material in thermal equilibrium, (4) requires each chemical component to satisfy $\mathbf{J} = 0$. It is the above thermodynamic analysis that produces the new conclusion of this paper, namely:

**Revised Conclusion #1** – Any material
(Requires charge transport equation {see (4)} and $\mathbf{J} = 0$)
The electric field $\mathbf{E}$ must be equal to zero *everywhere* inside a material in thermal equilibrium – except near an interface where an electric field must exist to counter the diffusion of free charges across the interfacial boundary. The current density on the other hand must be zero everywhere in order to satisfy the laws of thermodynamics.

Here again, this is a new way of looking at an interface, but it requires the recognition that an interface does not obey Ohm's law and that it is incorrect to apply the old Conclusion #1.

*D. Diffusion Potential - Conductors*

Based on the revised conclusions above it is a simple matter to determine the exact nature of the potential drop across an interface formed by two different conductors. For conductors only one charge dominates, namely free electrons. For two conductors in contact the interfacial electro-thermal equilibrium (IETE) criteria is given by Revised Conclusion #1 where (4) reduces to (2) and the electric field is created by the diffusion of free electrons across the interface, so the electric field can be defined as the diffusion induced field $\mathbf{E} = -\nabla \Psi$ where $\Psi$ is the potential induced due to diffusion. But since $\sigma = s\rho b$, where $b$ is the mobility of the free electrons, the solution to (2) for the thermal equilibrium condition $\mathbf{J} = 0$ is

$$\rho = \rho_0 e^{-\frac{sq}{kT}(\psi-\psi_0)} \qquad (5)$$

where the Einstein relation $qD = bkT$ was used and $s$ is the sign of the free charge, namely, $s = -1$ for free electrons [33]. If, in thermal equilibrium, an isolated conductor A has $\rho_{0A}$ free electrons (associated with its chemical component $N_A$) and if an isolated conductor B has $\rho_{0B}$ free electrons (associated with its chemical component $N_B$) then when the two conductors are brought together, each conductor must satisfy (5) and furthermore at the interface $\rho_{AI} = \rho_{BI}$ and $\Psi_{AI} = \Psi_{BI}$, from which (5) gives

$$\psi_{0A} - \psi_{0B} = -\frac{kT}{sq}\ln\frac{\rho_{0A}}{\rho_{0B}} \tag{6}$$

as the potential drop across the interface. Hence, it is clear that the IETE criteria require a potential difference across the interface whenever the concentrations of free electrons available between the two conductors differ.

Essentially, the IETE criteria given by (4) and $\mathbf{J} = 0$ will be involved no matter if the materials are solid, liquid or gas or are classified as conductor, semiconductor or insulator. However, each interface must be treated separately as all charged chemical components must be accounted for in the analysis. Furthermore, for liquids and gases not just electrons, but positive and negative ions must be accounted for in the analysis.

## VI. Conclusions

The concept of a surface having zero thickness came about more than 130 years ago when electricity was considered to be a weightless fluid and there was no knowledge of the existence of the electron and proton. As a result it was reasonable to many back then to assume that charges could be treated as point charges and be on a surface of zero thickness. When the laws of thermodynamics are applied to a conductor in thermal equilibrium, Ohm's law requires that the electric field inside the conductor be zero everywhere. If a conductor is charged the zero E-field requirement meant this charge must not be inside, but rather be on the surface, and point charges on a zero thickness surface fit nicely into this situation. However, when two conductors come together to form an interface a potential difference is found to exist across the interface which is not predicted by the zero E-field and zero thickness requirements. This paper reexamined the above analysis and found that charges have a finite thickness so they must occupy space either inside or outside the conductor. When two materials come together, if free charges are present in either one or both materials, diffusion of charge across the boundary can occur. The diffusion of charge near the interface eliminates the homogeneous material assumption so Ohm's law is no longer valid and the full equation of charge transport (3) must be use in place of Ohm's law. At an interface in thermal equilibrium, thermodynamics requires the absence of current, yet free electrons near the interface produce a diffusion current whenever the free elec-tron density between the two materials is not the same. This diffusion current produces a local electric field, which results in a conductive current that in thermal equilibrium just offsets the diffusion current so as to produce a zero net current. As a result of the local electric field across the interface, a potential difference exists across the interface. For two conductors this potential difference is proportional to the natural log of the ratio of the free electron densities of the two materials and is given by (6). In the more general case of any two materials the interfacial electro-thermal equilibrium (IETE) criteria are given by (4) and $\mathbf{J} = 0$, and this criteria always requires a potential difference across an interface whenever the free charges available in the two materials differ. It is suggested that this IETE criteria will be important to all who study electrical effects at interfaces. In the past functions such as the electric field have been defined as piecewise continuous across an interface with the interface defined as a boundary condition. The IETE criteria suggest that, when a potential difference occurs across an interface, the electric field is actually continuous across that interface.